# Dirac-like equation for two-level atom

N.V. Brazovskaja and V.Ye. Brazovsky

Altai State Technical University, Barnaul, Russia.
braz@agtu.secna.ru

**Abstract**. We construct Dirac-like equation for describing two-level atom interacting with resonant laser field.

## 1. Introduction

The behaviour of a two-level particle in an external field - interesting task, analysis allows which one to receive a number of precisely solved models. In a quantum electronics in a resonant case it is possible to take into account only two energy levels of a molecule or atom (hereinafter for a simplicity - atom) - at interaction with laser radiation [1]. From the methodical point of view the attractiveness of research of interaction of the system of two-level molecules with a resonance emission is encompassed by possibilities to construct visible, visual logical chain from the analysis of initial positions of the theory up to computed outcome of concrete effect observed experimentally.

The radiation-matter interaction currently used is based on some relevant approximations that are still well verified in the current experiments: Firstly, it is assumed that the dipole approximation holds, that is the wavelength of the radiation field being much larger than the atomic dimensions. Secondly, the rotating wave approximation (RWA) is always assumed, meaning by this that just near resonant terms are effective in describing the interaction between radiation and matter, these terms being also described as energy conserving. Thirdly, the intensity of laser irradiation. On the one hand it must be more then spontaneous emission, on the other hand it does not lead to energy-level splitting by the condition:

$$\frac{\mu E}{\hbar \Gamma} < 1 \, , \quad (1)$$

where $\mu$ is the matrix element of the dipole moment of the transition resonant with radiation, E is the external field of laser irradiation, $\Gamma$ is the halfwidth of the transition.

With an external field E(t), the Hamiltonian for complete system, atom plus quantized electromagnetic field E in the limit of long wave-lengths and in the rotating-wave approximation, can be written as follows [2]:

$$H = H_f + H_a + V \, . \quad (2)$$

where $H_f$ is field energy, $H_a$ describe the energy of atom as

$$H = \frac{p^2}{2m} + \hbar \omega_a \frac{1}{2}(1 + \sigma_z) \quad (3)$$

where $\hbar\omega$ is energy of the transition, perturbation is

$$V = -\mathbf{dE}, \quad (4)$$

Usually researches of two-level atoms fulfil with usage of a two-component wave function. Then the operator of dipole transition of atom **d** is defined by

$$\mathbf{d} = \mu \mathbf{s} \quad (5)$$

here **s** expressed in terms of the Pauli matrices:

$$\sigma_x = \begin{pmatrix} 0 & 1 \\ 1 & 0 \end{pmatrix}; \quad \sigma_y = \begin{pmatrix} 0 & -i \\ i & 0 \end{pmatrix}; \quad \sigma_z = \begin{pmatrix} 1 & 0 \\ 0 & -1 \end{pmatrix} \quad (6)$$

Let us consider this situation more carefully. First, expression of perturbation μ**s**E consist two vectors, and E – is the polar vector, but **s** is the axial vector. It multiplication lead to pseudoscalar, but energy must be scalar. Second, Dirac show years ago that one need four-component wave function to describe two-level system, not two-component. On the other hand, representation (1) - (6) allow to explain a lot of experimental dataset. What does it mean? One can suggest it may exist some solutions we lose using this representation.

In this work, we attempt to formulate the basic points of a mathematical scheme of quantum treatment of two-level atoms using Dirac-like equation and four-component representation of wave function.

## 2. Dirac-like equation

We write Dirac-like equation for a single atom as follows:

$$(-\frac{\hbar}{i}\frac{\partial}{\partial t} - c(\alpha p) + dE - \beta mc^2 - \beta^1 \hbar\omega)\Psi = 0 , \quad (7)$$

where used ordinary Dirac matrices: vector **α**, which can be written using Pauli matrices:

$$\alpha = \begin{pmatrix} 0 & \sigma \\ \sigma & 0 \end{pmatrix}, \quad (8)$$

and β as follows:

$$\beta = \begin{pmatrix} 1 & 0 & 0 & 0 \\ 0 & 1 & 0 & 0 \\ 0 & 0 & -1 & 0 \\ 0 & 0 & 0 & -1 \end{pmatrix}. \quad (9)$$

Our equation differ from Dirac equation by term

$$\beta^1 = \begin{pmatrix} 1 & 0 & 0 & 0 \\ 0 & 0 & 0 & 0 \\ 0 & 0 & -1 & 0 \\ 0 & 0 & 0 & 0 \end{pmatrix}, \quad (10)$$

which allow to absorb and emit photon energy $\hbar\omega$.

The next problem is to define the representation of dipole moment operator **d**. One can see, the definition

$$d = \mu \mathbf{S} , \quad (11)$$

where **S** is ordinary Dirac matrix:

$$\Sigma = \begin{pmatrix} \sigma & 0 \\ 0 & \sigma \end{pmatrix}, \quad (12)$$

by using nonrelativistic approximation lead to representation (1) - (6).

Is it all right? Let us pay attention to notice in previous section. This representation of dipole moment operator mean the perturbation has a form:

$$-\mu \mathbf{S} \mathbf{E} \qquad (13)$$

But **S** is the axial vector and such combination can not be used. The nature has not kept to us choice. We can use for electromagnetic field only two combinations: **SB** and **aE** (B is magnetic field). Therefore the only form of perturbation in our case is

$$V = -\mu \mathbf{a} \mathbf{E}, \qquad (14)$$

and representation of dipole moment operator

$$\mathbf{d} = \mu \mathbf{a}. \qquad (15)$$

Then Dirac-like equation comes to the form:

$$(-\frac{\hbar}{i}\frac{\partial}{\partial t} - c(\alpha p) + \mu(\alpha E) - \beta mc^2 - \beta^1 \hbar\omega)\Psi = 0 \qquad (16)$$

We know that the solution of the Dirac-like equation can be written as superposition of particle $\Psi_p$ and antiparticle $\Psi_a$ terms. As soon as we consider velocities of atoms negligible against velocity of light this solutions can be writtern as follows:

$$\Psi_p = \begin{pmatrix} \Psi_1 \\ \Psi_2 \\ 0 \\ 0 \end{pmatrix} e^{-i\frac{mc^2 t}{\hbar} + i\frac{pr}{\hbar}}, \qquad \Psi_a = \begin{pmatrix} 0 \\ 0 \\ \Psi_3 \\ \Psi_4 \end{pmatrix} e^{i\frac{mc^2 t}{\hbar} + i\frac{pr}{\hbar}}, \qquad (17)$$

Further we use a following mathematical device. For each function $\Psi_p$ and $\Psi_a$ separately we shall define following canonical trancformation:

$$\Psi = W^* \Psi^1, \quad H^1 = W^* H W,. \qquad (18)$$

where function W defined as follows:

$$W = e^{\pm \frac{mc^2 t}{\hbar}} \qquad (19)$$

Here plus-signs and minus are used for functions $\Psi_p$ and $\Psi_a$ accordingly. Then we obtain for wave functions:

$$\Psi_p = \begin{pmatrix} \Psi_1 \\ \Psi_2 \\ 0 \\ 0 \end{pmatrix} e^{+i\frac{pr}{\hbar}}, \qquad \Psi_a = \begin{pmatrix} 0 \\ 0 \\ \Psi_3 \\ \Psi_4 \end{pmatrix} e^{+i\frac{pr}{\hbar}}, \qquad (20)$$

and for equation:

$$(-\frac{\hbar}{i}\frac{\partial}{\partial t} - c(\alpha p) + \mu(\alpha E) - \beta^1 \hbar\omega)\Psi = 0. \qquad (21)$$

Here we discard indexes of transformated funtions.

As soon as obtained funtions $\Psi_p$ and $\Psi_a$ are the solutions of the last equation, there superposition also are the solution of one. Then the general solution of last equation can be written as follows:

$$\Psi = \begin{pmatrix} \Psi_1 \\ \Psi_2 \\ \Psi_3 \\ \Psi_4 \end{pmatrix} e^{+i\frac{pr}{\hbar}}. \quad (22)$$

## 3. Concluding remarks

In the theory of electron the transition to a positron is carried out with the help of a charge conjugation and the change of a sign of the time. In our case the change of the sign of a charge corresponds to change of the sign of electrical dipole moment, that corresponds to rotational displacement of the dipole on a corner $\pi$. But in the theory of radiations the phase change on $\pi$ corresponds to the replacement of the radiation process by absorption. Thus particle and antiparticle solutions of the equation (20) can be interpreted as radiant and absorptive states of an atom.

Let us assume that really we did not obtain, but constructed Dirac-like equation (21) with solution (22) for two-level atom. Our equation automatically includes atoms rotation by light. It allows to calculate dipole-dipole interaction using quantum rules [3].

In conclusion, we have shown that two-level atom in an external field can be described by Dirac-like equation and a four-component wave function. It can not be made two-component without any padding suggestions.